\documentclass[prb,twocolumn]{revtex4-1} 

\usepackage{comment}
\usepackage{amsmath} 
\usepackage{amssymb}

\usepackage{graphicx} 

\usepackage{hyperref}
\hypersetup{colorlinks=true,linkcolor=red,citecolor=blue,urlcolor=black,bookmarksdepth=subsubsection} 

\begin{document}

\title{An optical $n$-body gravitational lens analogy}
\author{Markus Selmke}
\affiliation{*Fraunhofer Institute for Applied Optics and Precision Engineering IOF, Albert-Einstein-Str.\ 7, 07745 Jena, Germany}
\email{markus.selmke@gmx.de}
\homepage{http://photonicsdesign.jimdo.com}
\date{\today}

\begin{abstract}
Raised menisci around small discs positioned to pull up a water-air interface provide a well controllable experimental setup capable of reproducing much of the rich phenomenology of gravitational lensing (or microlensing events) by $n$-body clusters. Results are shown for single, binary and triple mass lenses. The scheme represents a versatile testbench for the (astro)physics of general relativity's gravitational lens effects, including high multiplicity imaging of extended sources.
\end{abstract}

\maketitle 

\section{Introduction}
Gravitational lenses make for a fascinating and inspiring excursion in any optics class. They also confront students of a dedicated class on Einstein's general theory of relativity with a rich spectrum of associated phenomena. Accordingly, several introductions to the topic are available, just as there are comprehensive books on the full spectrum of observations (for a good selection of articles and books the reader is referred to a recent Resource Letter in this Journal \cite{Treu2012}). Instead of attempting any further introduction to the topic, which the author would certainly not be qualified to do in the first place, the article proposes a new optical analogy in this field.

Gravitational lenses found their optical refraction analogy at least as early as 1969.\cite{Liebes1969} Over the following 50 years, several variants of logarithmic axicon lenses or other glass/plastic lenses with similar profiles (the simplest one being a wine glass stem) have been used to simulate and tangibly convey the effects of gravitational lensing in class rooms,\cite{Icke1980,Higbie1981,Higbie1983,Surdej1993,Adler1995,Nandor1996,Lohre1996,Huwe2015} physics labs\cite{Brown} or museums\cite{AMNH} around the globe. All of these simulators are for \textit{single masses or mass distributions only}, ruling out for instance access to the growing field of microlensing (an important tool in the search for exoplanets) \cite{Giannini2013,Gaudi2012} and exotic gravitational lenses.\cite{deXivry2009} Quoting the pioneers of the theory of binary lens systems, \cite{Schneider1986} one may call attention to 
\begin{quote}
[\dots ] the fact that most stars are members of a binary (or multiple) system and that galaxies appear in pairs, groups or clusters [\dots].
\end{quote}

Inspired by earlier studies on the caustics of floating bubbles,\cite{Selmke2019} the herein proposed model system thus uses liquid menisci around fixed solid discs to extend past physical models to \textit{multiple component} ($n$-body) \textit{gravitational lenses}. Although a given component's liquid meniscus shape (exponential in its large-distance fall-off) does not provide a perfect approximation to the logarithmic profile needed to generate a proper optical analogy to a point mass, nor to other commonly assumed mass distributions \cite{Surdej1993,Adler1995} (though resembling some models of exponential spiral galaxy disks \cite{Freeman1970}), the phenomenology is again very similar. Using thin (compared to the capillary length) vertical rods instead can even provide a logarithmic profile accurately mimicking point masses at the cost of a more difficult observation due to the smaller size of such an installation,\cite{Selmke2019} which is why larger discs were used in this work. While it appears as if constellations of two glass lens models may have been used in the past to mimic binary lenses \cite{WebLink}, the present model provides a \textit{perfectly smooth} and \textit{connected single} adjustable \textit{refracting interface geometry} (in a single plane) by the action of surface tension, also enabling the simulation of dynamic microlensing events.\cite{Treu2010}


\begin{figure}[bt]
\begin{center}\includegraphics [width=1.0\columnwidth]{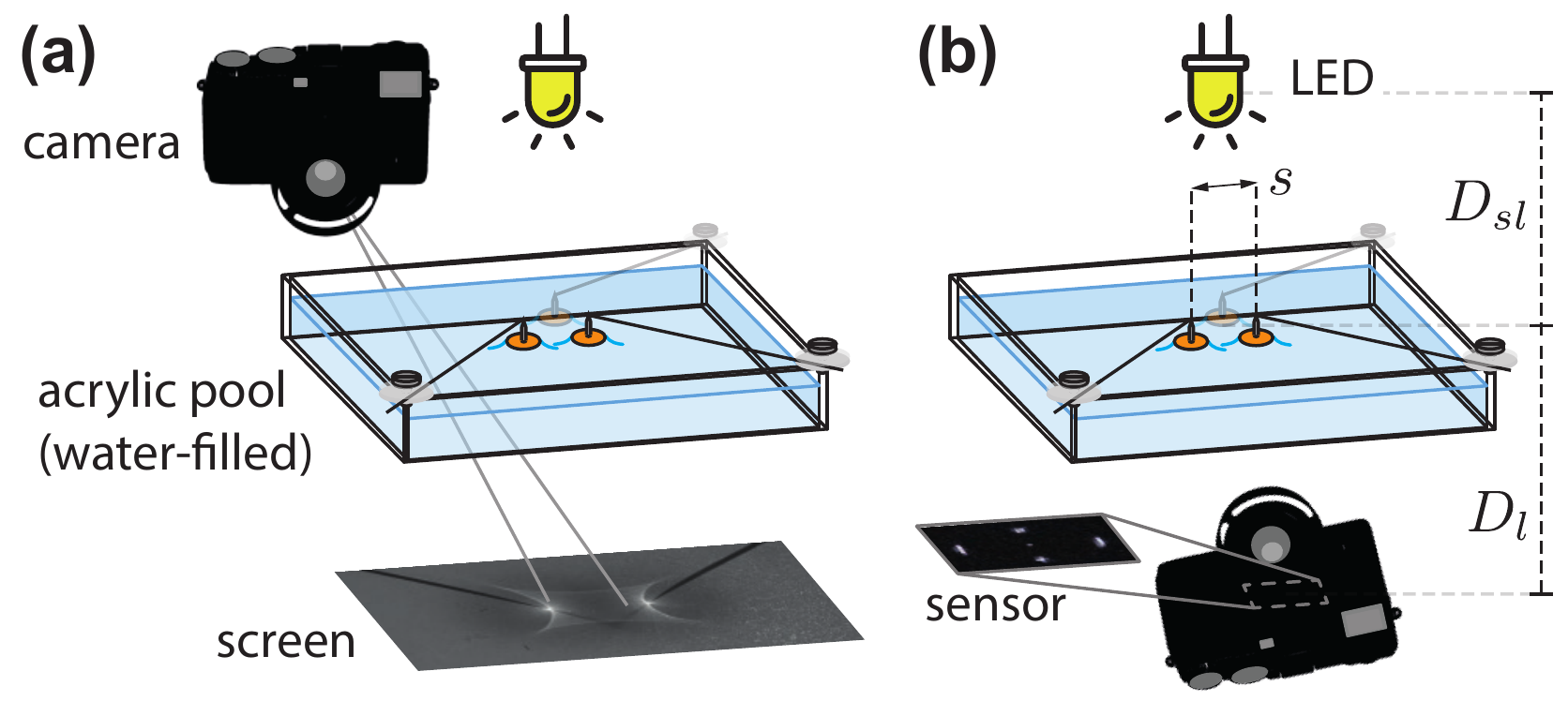}\end{center} %
\caption{A suitably mounted acrylic water basin \cite{Selmke2019} allows for views through menisci-cluster lenses. Drawing pins are held above the unperturbed water level by carbon fiber rods (diameter $\varnothing=0.5\,\rm mm$, pins: $\varnothing=10\,\rm mm$). The rods are attached adjustable to the basin using M3-screws and two plastic washers. Variants: \textbf{(a)} to study caustics, \textbf{(b)} to study images (by eye or camera). The distances were: $D_{l}=1.0\,\rm m$, $D_{sl}=1.3\,\rm m$, $s=1.4\,\rm cm$. Pictures were taken with a Fuji X-Pro 2 camera, XF60mm F2.4R Macro lens and large $f$-stop ${\geq 16}$.\label{Fig_Setup}} 
\end{figure}

To produce such strong gravitational $n$-body lensing\cite{Treu2010} or microlensing \cite{Giannini2013,Gaudi2012} analogies one may attach $n$ solid discs to support rods, e.g.\ via drawing pins (tacks) superglued to thin carbon rods, and position them in such a way that they \textit{pull up the liquid} around them, as shown schematically in Fig.\ \ref{Fig_Setup} (menisci sketched in blue, tacks in orange). The liquid (e.g.\ water) should be held in a flat-based basin and mounted such as to allow views through it (loc.\ Fig.\ 16(a) of Ref.\ [\!\!\citenum{Selmke2019}] shows an example of a wooden mount). Alternatively, separated bubbles (e.g.\ in the process of merger due to their mutual attraction \cite{Vella2005}) also produce in a dynamic way the same interface geometry, much in contrast to typical floating objects depressing water around them \cite{Lock2015}. While the bubble scenario has the advantage of an unobstructed geometry (no rods), the more controlled setup using positioned discs lends itself naturally for the proposed task and is considered in this article. 

More mathematically, in the gentle slope approximation the liquid surface perturbation $z=f(x,y)$ is governed by the linearized Young-Laplace equation\cite{Lock2015}
\begin{equation}
\nabla^2 f = a^{-2} f,
\end{equation}
with $a$ being the capillary length of the liquid (water: $a=2.73\,\rm mm$), and solved with proper boundary conditions. Light is then deflected via refraction at the interface $f$ by an angle (direction from incident to deflected ray)
\begin{equation}
\vec{\alpha}=(n-1)\nabla_{\rm 2D} f,
\end{equation}
where $\nabla_{\rm 2D}$ is the gradient operator acting on the $xy$-plane (see Appendix \ref{Appendix}). Both of these expressions resemble the case of gravitational lensing where the deflecting gravitational potential $\Phi$ is governed by Poisson's equation $\nabla^2 \Phi = 4\pi G\rho$ and $\vec{\alpha}=-(2/c^2)\int_S^O \nabla_{\perp} \Phi\,\mathrm{d}z$ (final tangent vector at observer $O$ minus initial tangent vector at source $S$. Typically, the deflection is defined vice versa.), where $\nabla_{\perp}$ (gradient perpendicular to the light path) can be approximated as the gradient operator $\nabla_{\rm 2D}$ acting on the plane of the projected thin lens $\phi=\int \Phi \mathrm{d}z$ \cite{Petters2001book}. The difference in signs in the deflection formulas is compensated for by the inverted topology of $\phi$ vs.\ $f$. Loosely then, the analogy is $-f\leftrightarrow \phi$ and $h \leftrightarrow \rho$. 

In contrast to the glass lens model \cite{Liebes1969,Icke1980,Higbie1981,Higbie1983,Surdej1993,Adler1995,Lohre1996,Huwe2015} this system, through the differential equation determining $f$ (also an elliptic PDE, a Helmholtz equation with imaginary wave number), provides an \textit{interactive analogy to the mass-warped space-time fabric} as well, somewhat like the rubber membrane visualization (with all its limitations \cite{Price2016}).


\section{Experimental Setup}
The experiment uses illumination by a distant conventional whitish LED light source. Though small in size, it still actually represents an extended source. However, for convenience, it is referred to as a "point light source" mostly in the manuscript as it is small enough to allow counting of non-overlapping images. Now, the experiment can be performed in at least two modi:

\begin{enumerate}
\item observation of the \textit{caustics} associated with a given $n$-body realization, see Fig.\ \ref{Fig_Setup}(a), or 
\item observation of the \textit{images} associated with a given $n$-body realization, see Fig.\ \ref{Fig_Setup}(b). 
\end{enumerate}

Experiments of type (a) are conveniently carried out first, i.e.\ prior to any effort to recreate a given astrophysical lensing image scenario as the caustic patterns are commonly used for classification of relevant configurations. They are typically shown in journal articles and can thus be used as a starting point to recreate a given analog configuration. Experiments of this type can also be used to recreate and study light-curves, that is time series of intensities for a given astronomical microlensing event as it would be observed for the integrated (unresolved) signal recorded with a telescope: To this end, one would photograph a certain caustic pattern and take a line profile through it in some direction. Alternatively, a dynamic caustic may be created (via a moving tack) with a fixed detector placed below the setup taking a time-series of intensities. The sequence of peaks appearing is characteristic of a certain $n$-body scenario.

Experiments of type (b) in contrast reveal the phenomenology of source images, akin to astronomical background objects lensed by heavy foreground objects as have for instance been catalogued in many surveys,\cite{Treu2010,Treu2012} including several exotic examples.\cite{deXivry2009} The multiplicity for various lensing configurations can be explored by counting the number of images observed for the light source as recorded by a camera or seen by the unaided eye when viewed through the setup. Here, the LED can be replaced by a printed image of any extended model source as well to study the complex distortion patterns.\cite{Icke1980,Wambsganss2010,Koopmanns2004} 

A fixed number of images corresponds to a given region of the caustic bound by folds and cusps in which the camera (or the eye of the observer) is placed. Crossing folds causes a change of the number of images by at least $\pm 2$ and by at most $\pm 4$.\cite{Blandford1986,Petters2001book,Nye1999Book} While admittedly the former correlation between caustic regions and image multiplicity is hard to make in the simplified setup considered here (cf.\ Fig.\ \ref{Fig_Setup}), the more controlled though light-sacrificing setup variant using a pinhole screen could be adopted for this purpose.\cite{Surdej1993} Effectively, setting the camera lenses' $f$-stop to high values functions as a pinhole in the present setup, though projecting on a smaller screen (the camera-chip).


\section{Three scenarios}
The complexity of pure $n$-point mass lensing scenarios (no added external shear) grows rapidly with increasing number $n$ of lenses. A thorough analysis and comprehensive description has been done for two-mass lenses.\cite{Schneider1986,Dominik1999,Erdl1993,Witt1993,Cassan2008} This situation is particularly useful for the search of exoplanets.\cite{Gaudi2012,Giannini2013} For three-mass lenses, extensive studies exist as well, although the system is already hardly tractable.\cite{Danek2015,Danek2019} Thus, only scenarios up to $n=3$ have been considered experimentally here.

For a single point mass ($n=1$), there are exactly $2$ images of a point source. For an extended lensing mass distribution (a \textit{non-singular} bounded transparent lens, e.g.\ an elliptical lens) the number of images is odd \cite{Burke1981}, and the observable image multiplicity depends on whether the central lensing mass is transparent or outshining a central image \cite{Evans2001} (whereby typically either $2$ or $4$ images have been observed). The number of images for $n>1$ point masses ranges from $n+1$ to $5n-5$ in increments of two,\cite{Rhie2003,Khavinson2006} i.e.\ for binary systems ($n=2$) from $3$ to $5$, and for triplet systems ($n=3$) from $4$ to $10$. For non-singular (extended) multi-component lenses, the odd-number image theorem \cite{Burke1981} holds as well \cite{Schneider1985}. Again, apparent discrepancies between observations and theory can typically be explained by unseen "ghost images" hidden by the deflectors \cite{Blandford1986,Giannoni1999} or strongly demagnified and thereby weak sub-images.\cite{Chang1984}

\subsection{Single mass gravitational lenses ($n=1$)}
\begin{figure}[bt]
\begin{center}\includegraphics [width=1.0\columnwidth]{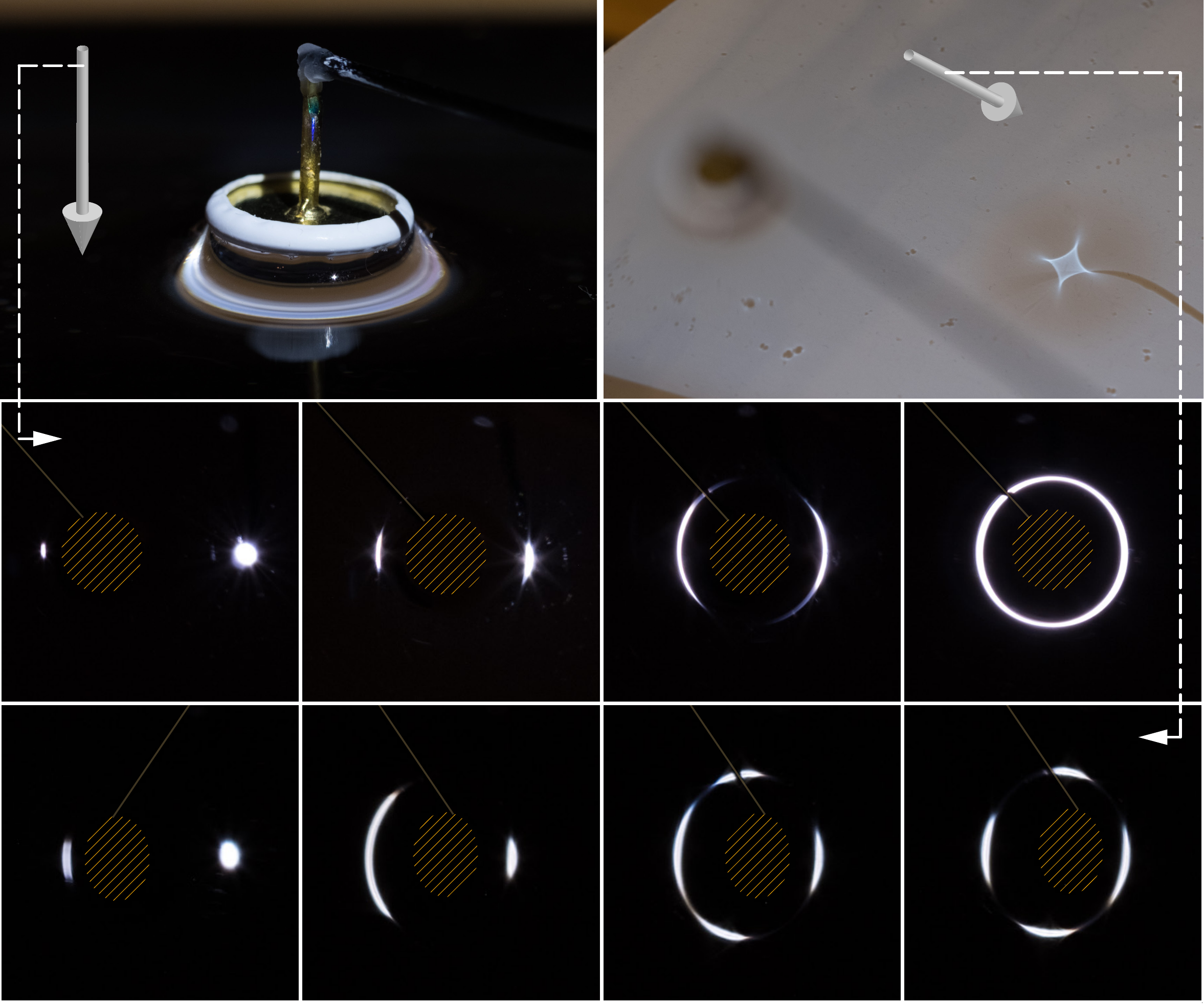}\end{center} %
\caption{Images for a \textit{single} lens using the setup of type (b). The top series was taken looking vertically through the lens towards the LED, whereas the bottom row corresponds to looking towards the LED under an angle of about $40^{\circ}$ to the vertical. In the top row, the image positions qualitatively agree with the series for gravitational lensing of an extended source by a point mass (cf.\ loc.\ {Fig.\ 8.20(a)-(c)} in [\!\!\citenum{Nye1999Book}], including an Einstein ring). The bottom row image positions correspond to gravitational lensing of an elliptical mass distribution (or a point mass with external shear, cf.\ loc.\ {Fig.\ 8.20(d)-(f)} in [\!\!\citenum{Nye1999Book}], including asymmetric Einstein cross \cite{Blandford1989}).\label{Fig_SingleLens}}
\end{figure}
The best known example of strong gravitational lensing is that of an extended source by a single mass in the most simple scenario: A configuration in which the observer, the lensing mass (at least axis-symmetric) as well as the lensed source are lying on a single axis: in this case, an \textit{Einstein ring} is observed \cite{Blandford1989,Nye1999Book,Treu2010} (e.g.\ SDP.81, LRG 3-757, B1938+666). For the case of a single tack of radius $R$, setting $\left.f(r)\right|_{r=R}=h$, the angular Einstein ring radius $\theta_E(D_l,D_{sl})$ may be computed from the approximate profile $f=h\exp(-(r-R)/a)$ and the experimental configuration, yielding an expression indeed resembling the gravitational lens case (see Appendix \ref{Appendix}. In the experiments, $\theta_E\sim 1^{\circ}$). Now, when either the lens or the source are displaced away from the axis, a splitting in two elongated arcs is seen for an extended source, or a splitting into two points for a point source (one inside, a brighter one outside the Einstein ring radius). For larger displacements, one of the images becomes significantly fainter while the other becomes unlensed. The experimental recreation using a single positioned tack is shown in the central row of Fig.\ \ref{Fig_SingleLens}.

When the single lensing mass is elliptical (as projected in the lens plane), the symmetry is reduced and the Einstein-ring for the perfect alignment scenario breaks up into the so-called \textit{Einstein cross} \cite{Blandford1989,Nye1999Book,Treu2010} (e.g.\ QSO 2237+0305, J2211-0350). For any displacement of either the source or the lens, two of the images making up the cross move towards one of the remaining two images before finally merging to yield then only two images. Upon further displacement, again one one of the images becomes fainter while the other becomes the unlensed one. The experimental recreation is shown in the bottom row of Fig.\ \ref{Fig_SingleLens}, where an \textit{inclined view} through the setup gives an elliptical lens as projected perpendicular to the viewing axis. Note, that a central image is obstructed by the opaque disc, such that the observed numbers of images ($2$ and $4$) are consistent with the theoretical expectations for an extended asymmetric mass distribution. A transparent acrylic disc instead of a tack could possibly overcome this shortcoming of the model. %

The same phenomenology also occurs for a point source with external shear (a Chang-Refsdal lens),\cite{Chang1984,Blandford1989} mimicking an asymmetric environment (say galaxies or clusters near the lenses, or structures along the ray path). External shear may be thought of as an extremely asymmetric two-mass lens system,\cite{Schneider1986,Dominik1999} yielding $3$ and $5$ images, with the central ones unobserved (effectively totalling then the $2$ and $4$ images of the Chang-Refsdal lens \cite{Chang1984}).


\subsection{Two mass gravitational lenses ($n=2$)}
For two point masses, the phenomenology becomes richer. Three qualitatively different configurations can be identified for gravitational (micro)lensing by binary mass systems: wide, intermediate, and close, each having qualitatively different caustics.\cite{Schneider1986,Dominik1999,Erdl1993,Witt1993,Cassan2008} The categorization of a given system is determined by two parameters only, the mass ratio (in this analogy: the ratio of meniscus heights $h$) and the distance $s$ (cf.\ Fig.\ \ref{Fig_Setup}(b)) of the two involved masses (or tacks). The resulting caustics of two point masses entail only cusps and folds.

\begin{figure}[bth]
\begin{center}\includegraphics [width=1.0\columnwidth]{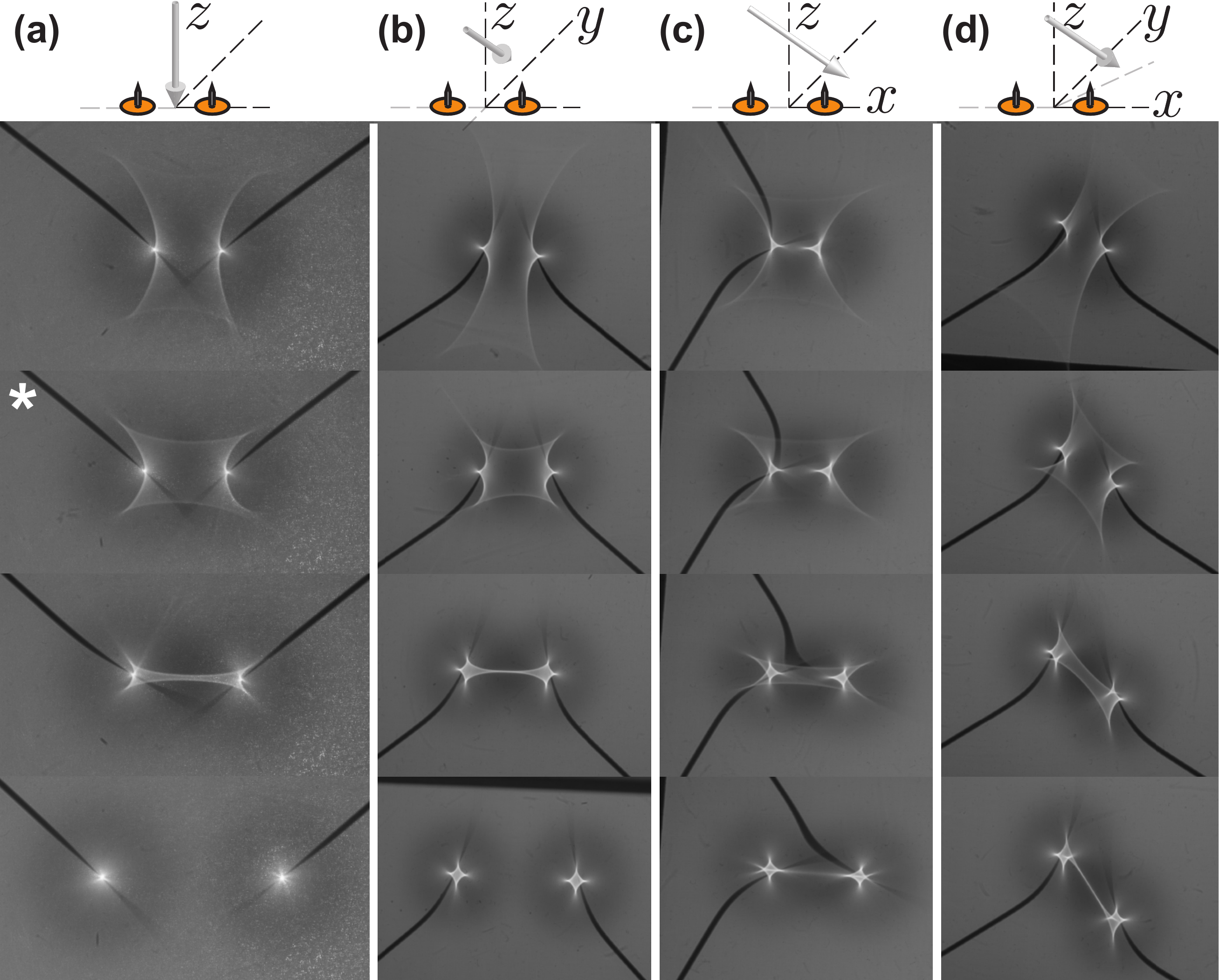}\end{center} %
\caption{Caustic metamorphoses (top to bottom) below two tacks for decreasing distances $s\sim (2.5 - 1)\,\rm cm$ (cf.\ Fig.\ \ref{Fig_Setup}, $s=2R=1\,\rm cm\equiv$ contact) and four different illumination scenarios: \textbf{(a)} Light source directly above, \textbf{(b)} inclined ($\Gamma\sim 50^{\circ}$) incidence perpendicular, \textbf{(c)} parallel and \textbf{(d)} diagonal to the symmetry axis. A configuration similar to the one marked with an asterisk (*) was chosen for the images in Fig.\ \ref{Fig_DoubleLens}.\label{Fig_Caustics}} 
\end{figure}

\begin{figure}[bt]
\begin{center}\includegraphics [width=1.0\columnwidth]{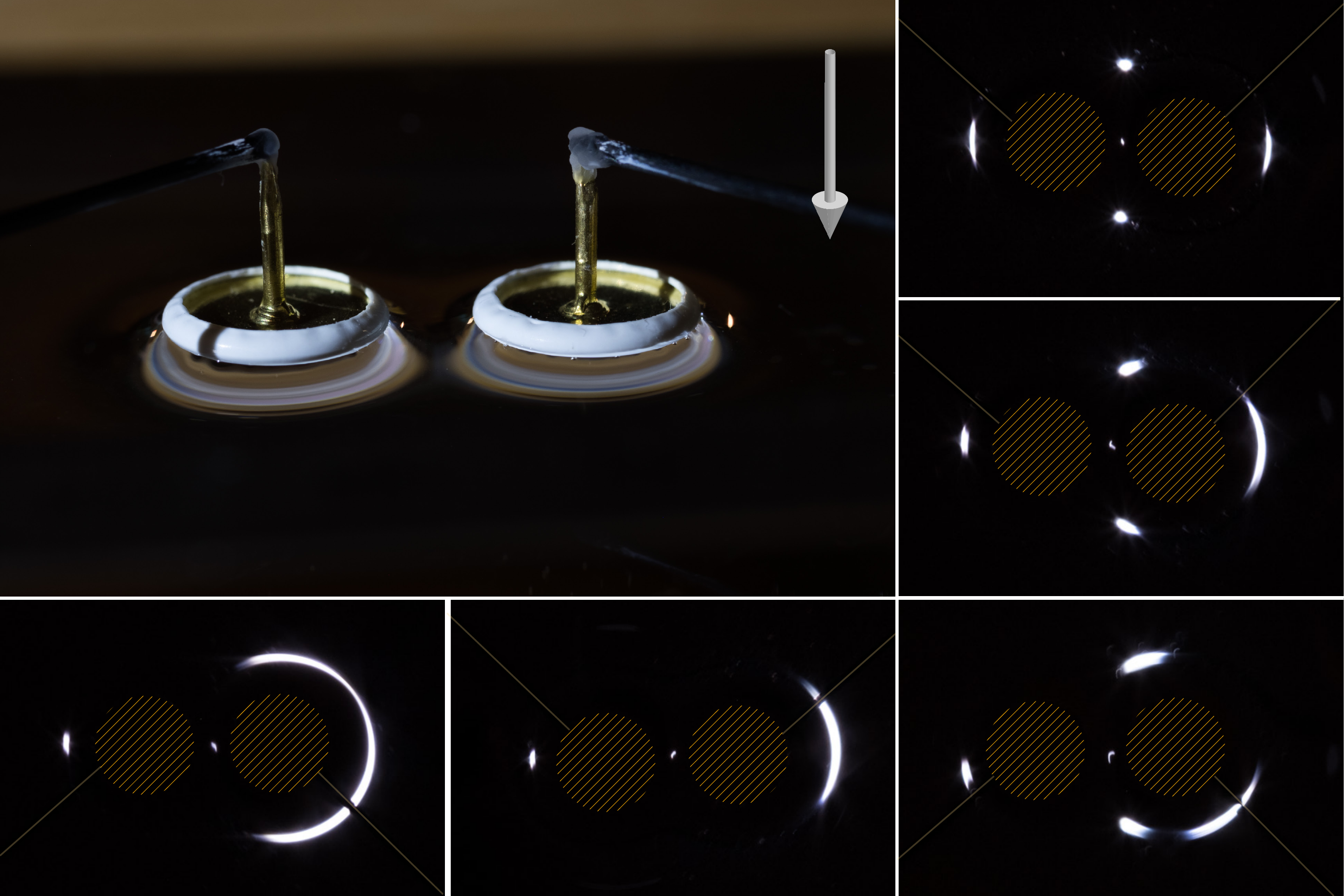}\end{center} %
\caption{Images for a \textit{double} (binary) lens using the setup of type (b). The image locations qualitatively agree with those shown in loc.\ Fig.\ 6 in [\!\!\citenum{Schneider1986}] for a gravitational two-point-mass lens for an intermediate distance case (corresponding caustic in loc.\ Fig.\ 2, likewise observed here (not shown)).\label{Fig_DoubleLens}}
\end{figure}

An experimental recreation of the no-shear system (vertical view) began by adjusting the two tacks to yield approximately the characteristic intermediate-distance caustic (type (a) experiment, see Fig.\ \ref{Fig_Caustics}). The resulting images for the corresponding type (b) experiments are shown in Fig.\ \ref{Fig_DoubleLens}, yielding from $3$ up to $5$ separate images of the single point source when including the central weak image (which has a high probability to be missing in astronomical observations). The image configurations attainable may be compared to astronomical examples such as CLASS B1608+656\cite{Koopmans1999} or SL2SJ1405+5502.\cite{deXivry2009}

It must be noted that the "close" configuration\cite{Schneider1986,Dominik1999,Erdl1993,Witt1993,Cassan2008} could not be reproduced at all, see Fig.\ \ref{Fig_Caustics}. Also, on close inspection it was found that for the two extended discs the 6-cusped intermediate caustic expected for point sources was easily perturbed by small misalignments: the two cusps on the symmetry axis then evolved into tiny \textit{butterfly caustics}. This caustic structure resembled the one of two point masses with added external shear, as shown in loc.\ Fig.\ 11(a) of [\!\!\citenum{Witt1993}] (rather than e.g.\ loc.\ Fig.\ 2 ($X=0.5$) of [\!\!\citenum{Schneider1986}]). Indeed, theory tells that with added external shear (effectively then corresponding to at least three lensing masses), or again by perturbing the spherical symmetry by an extended mass distribution (or elliptical ones, i.e.\ considering an inclined view through the setup), higher-order caustics can appear: the swallowtail and the butterfly caustics, with a correspondingly increased (odd) number of images \cite{Witt1993}. 

It should be a worthwhile advanced laboratory course exercise to try and access the higher multiplicities through highly inclined view experiments of type (b) corresponding to the caustics shown in the third column of Fig.\ \ref{Fig_Caustics} (symmetry axis connecting the two tacks in the plane of inclined incidence). 

%

\subsection{Three mass gravitational lenses ($n=3$)}
For three point masses, the phenomenology becomes even richer \cite{Danek2015,Danek2019}. Already, there are five parameters required to fully describe the situation: two mass ratios (menisci height ratios) along with three relative positions defining the two-dimensional configuration of the masses (tacks). Many gravitational triple lens scenarios have been observed, although almost all have been very asymmetrical in nature: either binary stars with a single planet or single stars with two planets \cite{Gaudi2012,Danek2015,Danek2019}. These situations can be viewed as perturbed single mass or binary systems. To the best of the author's knowledge, the most prominent triple mass system where all three masses contributed roughly equally to the lensing has been identified in 2001: CLASS B1359+154 is a group of three compact galaxies lensing a radio source (and its host galaxy), resulting in 6 images (likely a scenario involving extended lens components yielding $9$ images, $6$ of which are observable) \cite{Rusin2015}. 

\begin{figure}[bt]
\begin{center}\includegraphics [width=1.0\columnwidth]{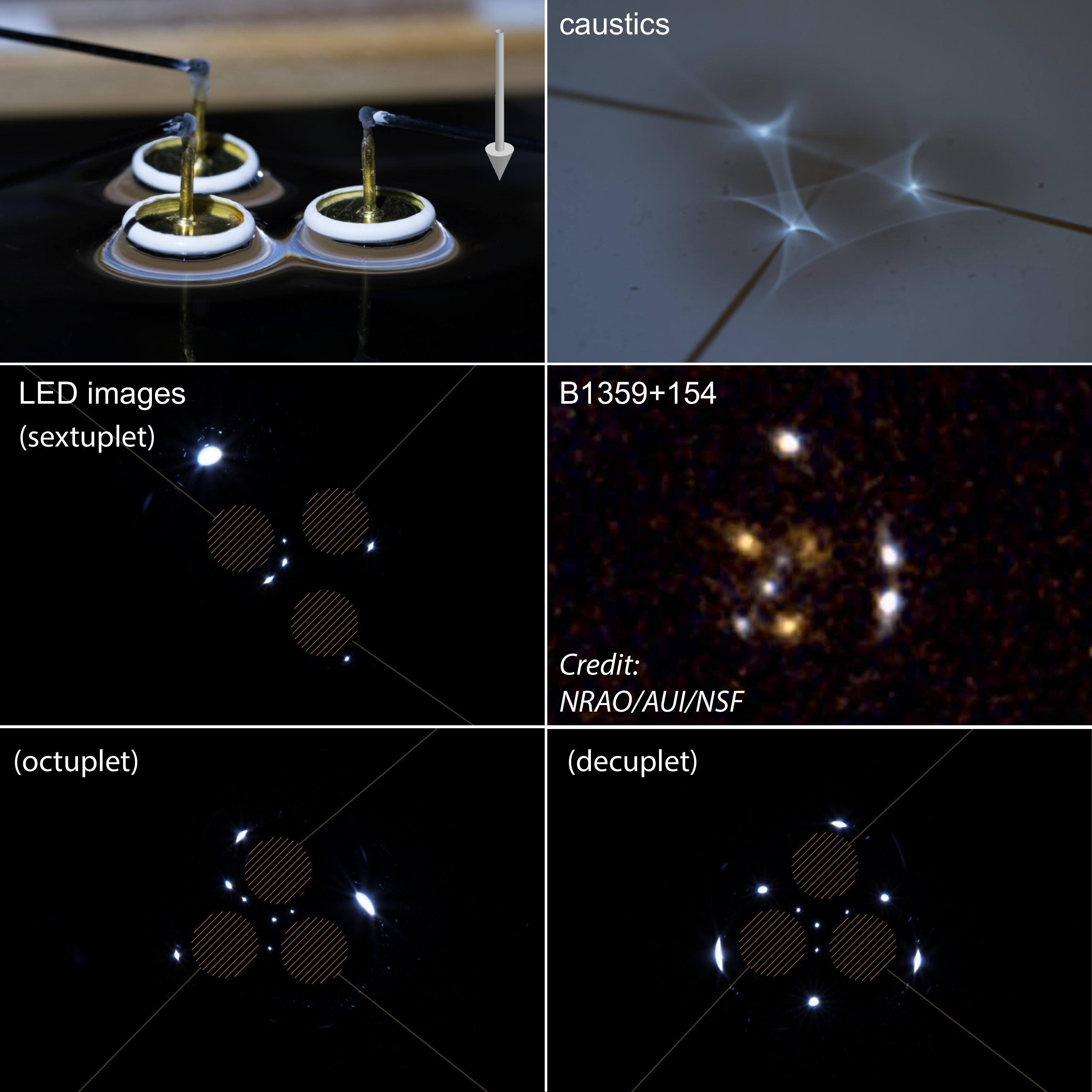}\end{center} %
\caption{Images for a \textit{triple} lens using the setup of type (b). The top and center rows show a slightly asymmetric configuration of the lenses, and the image locations qualitatively agree with those observed for CLASS B1359+154 \cite{Rusin2015}. The bottom row shows a symmetric configuration in which the largest multiplicity which could be observed was $10$.\label{Fig_TripleLens}}
\end{figure}

An experimental recreation using three positioned tacks (roughly attempting to recreate the asymmetric 3-mass configuration of the astronomical counterpart) is shown in Fig.\ \ref{Fig_TripleLens}. A fair match was found rather quickly, where the closeness was somewhat surprising and is certainly to some extend accidental, given that no optimization of the menisci heights, their diameters or their detailed positions was undertaken and given that the analogy is rather to a system of point masses. The matching image was one of several found image configurations, where up to 10 images were observed (in line with the expectation for a three point mass system).

It should be an interesting advanced laboratory course exercise to try to recreate the zoo of caustics reported for various different parameter triple lens systems \cite{Danek2015}.
%

\section{Alternative: free-forms optics?}
\begin{figure}[bt]
\begin{center}\includegraphics [width=1.0\columnwidth]{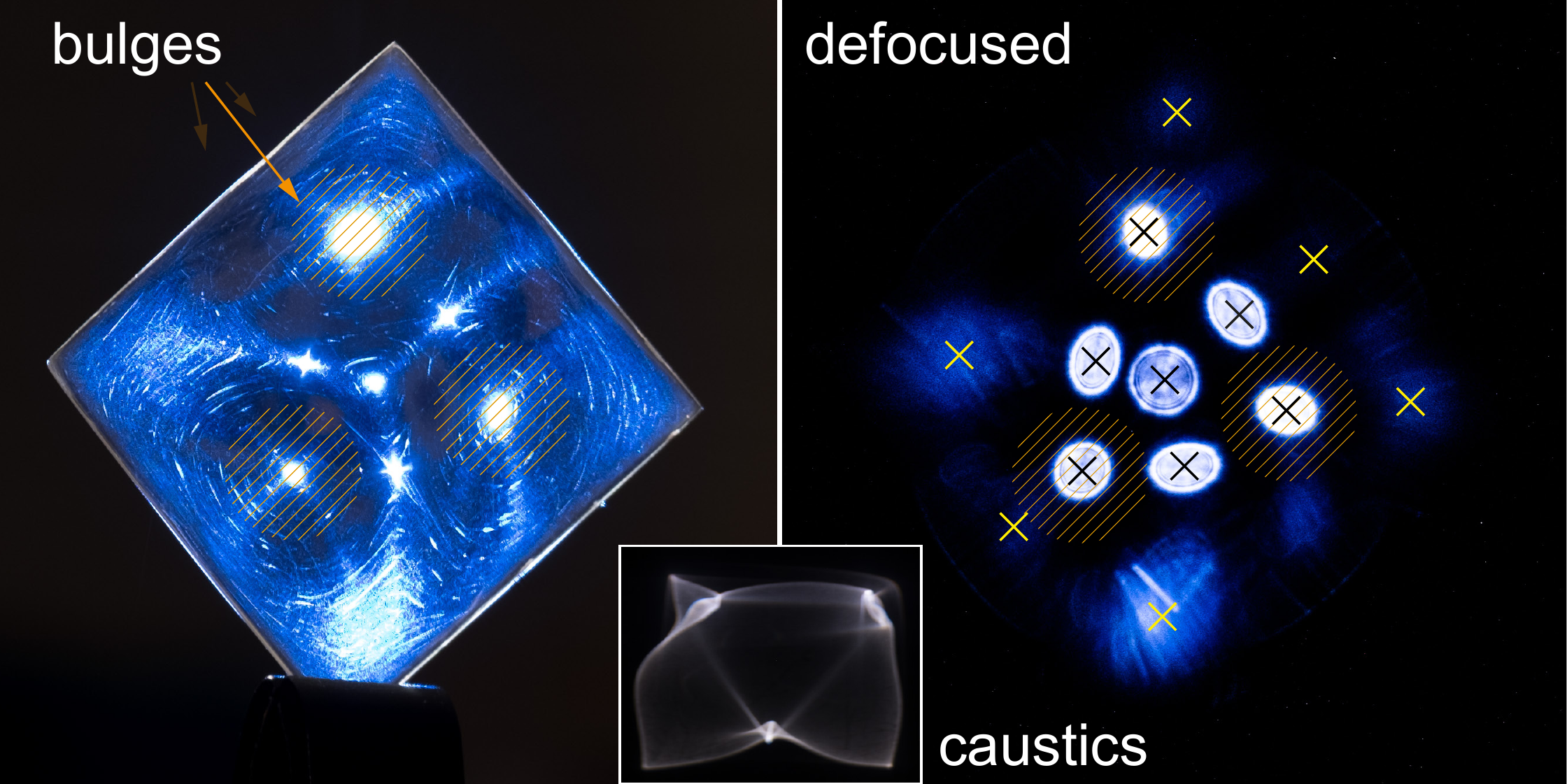}\end{center} %
\caption{(Failed) attempt of a hand-crafted freeform acrylic triple lens. $7$ images are clearly visible (defocused bright spots, black crosses), while indications of 6 more can only be guessed (weak and deformed spots, yellow crosses). The $3$ images due to the central bulges (locations marked by orange discs) are absent in the tack menisci model (cf.\ Fig.\ \ref{Fig_TripleLens}).\label{Fig_TripleLensAcrylic}}
\end{figure}

The present model's strengths can also be appreciated from the difficulty one encounters when trying to create a freeform multi-component lens mimicking the phenomenology of the triple lens by hand. Such an attempt is shown in Fig.\ \ref{Fig_TripleLensAcrylic}, which was done starting from a $5\,{\rm cm}\times 5\, {\rm cm} \times 0.5\, {\rm cm}$ acrylic sheet working with a rotary tool and grinding papers of different grits. The caustics (inset) were fairly chaotic around three highly aberrated foci, whereas the images observed were incomplete and very much imperfect. The missing images are due to missing curvature at the periphery of the acrylic sheet where instead a monotonic decay was apparently imposed. Using 3D-printed negative molds and UV-curable resin should allow a better match with expectations, though the resulting piece would still be static! It is the menisci model's strength to provide a readily adjustable and possibly dynamic way to realize a perfectly smooth $n$-component lensing geometry which is automatically decaying for large distances and interpolating in-between components to yield all images expected from the astrophysical counterparts.

\section{Conclusion}
Using the simple experimental scheme proposed in this article its potential has been demonstrated using the three cases of single, double and triple mass systems. In each case, the phenomenology of the associated caustics as well as the image multiplicity and image configurations could be reproduced faithfully for selected configurations. Admittedly, the analogy has not been probed exhaustively, i.e.\ the immense parameter spaces for $n=2,3$-body lenses have not been sampled wholly (including mass / menisci height ratios), and neither have the effects of distances nor transitions from weak to strong lensing etc.\ been investigated. Still, the utility of the model was hopefully sufficiently motivated and leaves much room for experimentation to still be done (cf.\ also the "Chashire Cat Challenge" in Appendix \ref{Chashire}). The present manuscript also leaves room for a deeper theoretical exploration of the optical analogy.

In summary, the system affords optically smooth surfaces representing analog scenarios which closely mimic ideal theoretical predictions for the strong lensing by $n$ point masses. Although the underlying details of the imaging are different, the analogy is close enough to afford a good match of the imaging characteristics. The setup should thus be a valuable advanced demonstration piece, supplementing the simpler single mass (distribution) glass simulators,\cite{Liebes1969,Icke1980,Higbie1981,Higbie1983,Surdej1993,Adler1995,Nandor1996,Lohre1996,Huwe2015} and maybe also a tool allowing to gain insight into more complex gravitational lensing configurations by hinting at an analogy to the remote field of optical caustics of swimming objects.

\section{Acknowledgements}
I thank K.-H. Lotze for fruitful discussions on glass lens experiments, which prompted the attempted free-form triple lens.

\newpage
\section{Appendix}\label{Appendix}
Since the differential equation for the surface, $\nabla^2 f = a^{-2} f$, is linear, solutions for the individual discs may be superimposed, just like in the gravitational lensing case. In cylindrical coordinates, and for a single disc only, the differential equation has the solution $f(r)=h K_0(r/a)/K_0(R/a)$, with $K_0$ being the modified Bessel function of the second kind of zeroth order, and $R$ being the radius of the disc. 

\begin{figure}[bth]
\begin{center}\includegraphics [width=1.0\columnwidth]{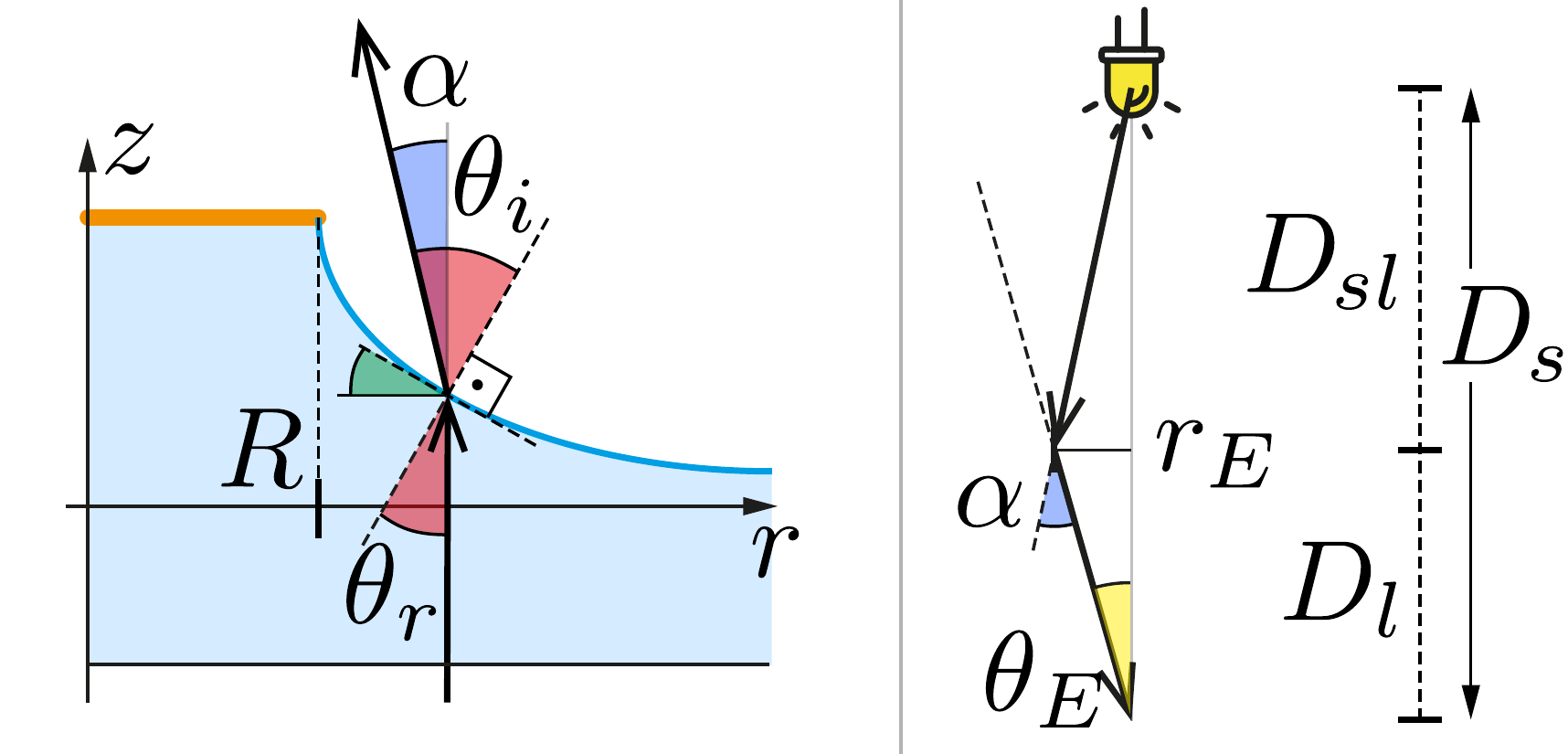}\end{center} %
\caption{Calculation of the deflection angle $\alpha=|\vec{\alpha}|$ (blue) by considering the inverse situation (light from below): Snell's law for small angles (red) reads $n\theta_r=\theta_i$, also $\theta_r=-f'$ (green) such that $\alpha=\theta_i-\theta_r\approx -f' (n-1)$. The Einstein ring radius $r_E$ then follows (see right part): $r_E/D_l + r_E/D_{sl}=\alpha(r_E)$, and the angular diameter is $\theta_E \approx r_E/D_{l}$. \label{Fig_Sketchf}} 
\end{figure}

From Fig.\ \ref{Fig_Sketchf} the the bending angle vector can be seen to be $\vec{\alpha}=(n-1)\nabla f$ (where $\nabla f\propto -\mathbf{\hat{\rho}}$ for raised menisci points towards the optical axis), where in the gentle slope approximation $\nabla=\mathbf{\hat{\rho}}\partial / \partial \rho + \dots$ in cylindrical coordinates may be taken as the gradient operator $\nabla_{\rm 2D}=\mathbf{\hat{x}}\partial / \partial x + \mathbf{\hat{y}}\partial / \partial y$ acting in the $xy$-plane.

For the parameters used in the experiments (${h\sim 5\,\rm mm}$), and using the geometry of the lensing situation as depicted in Fig.\ \ref{Fig_Sketchf}, the Einstein angle can be found numerically by solving
\begin{equation}
r_E/D_l + r_E/D_{sl}=\alpha(r_E)
\end{equation}
to find $\theta_E \sim 0.7^{\circ}$ (angular diameter of ${\sim 1.5^{\circ}}$). This angular diameter is indeed small compared to the (lateral) angle of view of $22.3^{\circ}$ of the used macro camera lens and corresponds nicely to the observed ca.\ $7.5$ percent lateral filling fraction of the APS-C sensor (the images in the Figures were cropped)

Using instead the exponentially damped approximation \cite{Lock2015} $f(r,\phi)=h \exp(-(r-R)/a)$ for a single lens, yields the following \textit{analytical} expression for the radius $r_E$ of the Einstein ring:
\begin{equation}
r_E\approx a \cdot W\left(\frac{(n-1) h\exp(R/a)}{a^2}\frac{D_l D_{sl}}{D_s}\right)\label{EqrE}
\end{equation} 
where $D_s=D_l+D_{sl}$ was used (an identity which is not true on cosmological scales for gravitational lensing) and $W(x)$ is the Lambert W-Function (ProductLog-function), i.e.\ the inverse of $x\exp(x)$. The corresponding Einstein angular radius is $\theta_E=r_E/D_l$. 

Again using the parameters of the experiments, the result is $\theta_E\sim 0.8^{\circ}$ (an angular diameter of $\sim 1.6^{\circ}$). 

Expression \eqref{EqrE} is similar to the gravitational Einstein radius expression, which for $\alpha=4GM/r c^2$ solves to $r_E^2=(4 G M /c^2)\times D_l D_{sl}/D_s$, and $\theta_E=r_E/D_l$.

\section{The Chashire Cat Challenge}\label{Chashire}
The "Cheshire Cat" system should be a worthwhile example of a beautiful and well-known \textit{wide binary} lens\cite{Shin2008}, resembling the face of a smiling cat. It is a complex system consisting of multiple arcs on two different Einstein radii, foreground and lensing galaxies. Luckily, the hard work of figuring out the system's likely configuration has already been done using spectroscopy and gravitational lens modeling\cite{Belokurov2009}: At least $7$ images can be clearly identified and belong to \textit{two sources in different planes} ($D_{s,1}\ne D_{s,2}$) (take two LEDs at different distances), whereas the two lensing galaxies lie roughly in the same plane at $D_l$ (use two tacks). A third foreground galaxy not partaking in lensing forms the nose of the cat (add a third LED between the observer and the lenses). By trial and error, given the information derived by the astronomers, a recreation of the Cheshire Cat should be possible for an ambitious experimentalist or within an advanced lab course. The cat's eyes could be superposed images of the tacks (or use phosphorescent paint on the two tacks).

\end{document}